\documentclass[aps,prl,twocolumn,showpacs,amsmath,amssymb,amsfonts,sections]{revtex4}

\usepackage{graphicx}

\begin{document}


\title{Gravitational and electromagnetic fields near a de~Sitter-like infinity}

\author{Pavel Krtou\v{s}}

\author{Ji\v{r}\'{\i} Podolsk\'y}

\author{Ji\v{r}\'{\i} Bi\v{c}\'ak}

\affiliation{
  Institute of Theoretical Physics,
  Faculty of Mathematics and Physics, Charles University in Prague,\\
  V Hole\v{s}ovi\v{c}k\'{a}ch 2, 180 00 Prague 8, Czech Republic
  }

\date{May 30, 2003}        

\begin{abstract}
We present a characterization of general gravitational and electromagnetic
fields near de~Sitter-like conformal infinity which supplements the standard 
peeling behavior. This is based on an explicit evaluation of the dependence 
of the radiative component of the fields on the null direction
from which infinity is approached.  It is shown that the directional pattern of radiation 
has a universal character that  is determined by the  algebraic (Petrov) type of the spacetime.
Specifically, the radiation field vanishes along directions opposite to principal null directions. 
\end{abstract}


\maketitle


A direct observation of gravitational waves will be properly understood 
only when it can be compared with reliable predictions supplied by numerical relativity. 
To make such predictions is difficult since, among other things, no 
rigorous statements are available which relate the properties of sufficiently 
general strong sources to the radiation fields produced. 
Only a few explicit radiative solutions of Einstein's equations 
are known which can be used as test beds for numerical codes 
(see e.g. \cite{Bicak:Ehlers,BicakKrtous:NEBX}), in particular,
spacetimes representing \vague{uniformly accelerated particles or black holes}.
These have also been the main inspiration 
for our present analysis of the general asymptotic properties of radiation
in spacetimes with a positive cosmological constant $\Lambda$.
The motivation for considering de~Sitter-like universes arises 
not only by their role in inflationary theories 
but also by the fact that many \emph{non-vacuum} cosmological models 
with ${\Lambda > 0}$ (as suggested by recent observations) approach the
de~Sitter universe asymptotically in time (\vague{cosmic no-hair conjecture}) and hence
have a de~Sitter-like infinity. 

We have recently constructed the test fields of uniformly accelerated charges 
in de~Sitter spacetime (the Born solutions generalized for ${\Lambda>0}$) 
and investigated how their radiative properties depend on the way in which 
infinity is approached \cite{BicakKrtous:FUACS}. 
Somewhat surprisingly, we have found analogous results \cite{KrtousPodolsky:RABHDS}
in the case of the exact solution 
of the Einstein-Maxwell equations with ${\Lambda>0}$, namely the \vague{charged C-metric}
describing a pair of charged accelerated black holes in a de~Sitter universe.

In the following, we shall demonstrate that the directional pattern of radiation 
near an infinity of de~Sitter type has a universal character that is determined by the 
algebraic (Petrov) type of the spacetime. In spacetimes which are asymptotically 
de~Sitter, there are two disjoint past and future (conformal) infinities $\scri^-$  
and $\scri^+$, both  spacelike \cite{Penrose:1965,PenroseRindler:book}. 
In such spacetimes both  cosmological horizons and 
event horizons for geodesic observers occur and, consequently, 
advanced effects have to be present if the fields are smooth \cite{BicakKrtous:ASDS}. 
Curiously enough, with ${\Lambda>0}$ the global existence has been established
of asymptotically  simple vacuum solutions which differ on an arbitrary given Cauchy surface by a 
finite but sufficiently small amount from de~Sitter data  
\cite{Friedrich:1998}, while an analogous result for data close to Minkowski (${\Lambda = 0}$) 
is still under investigation \cite{Friedrich:1998,BicakKrtous:NEBX}. Thus, many vacuum asymptotically simple spacetimes 
with de~Sitter-like $\scri^+$ do exist. Assuming their existence, 
Penrose proved already in 1965  \cite{Penrose:1965,PenroseRindler:book} that both the gravitational 
and the electromagnetic fields satisfy the peeling-off property with respect 
to null geodesics reaching any point of $\scri^+$. This means that along a null geodesic 
parametrized by an affine parameter $\afp$, the part of any spin-$s$ 
zero rest-mass field proportional to $\afp^{-(k+1)}$, ${k = 0,1,\dots,2s}$,
has, in general, ${2s-k}$ coincident principal null directions. 
In particular, the part of the field that falls off as $\afp^{-1}$ is a 
\emph{radiation} (\vague{null}) \emph{field}. The peeling-off property is easier to 
prove with ${\Lambda>0}$ than in asymptotically Minkowskian spacetimes when $\scri^+$ 
is a null hypersurface \cite{Penrose:1965}. With a spacelike $\scri^+$, however, one can 
approach any point on $\scri^+$ from infinitely many different null directions and, 
consequently, the radiation field becomes mixed up with other components of the field 
when the null geodesic is changed. This fact of the \vague{origin dependence} of the 
radiation field in case of a spacelike $\scri^+$ has been repeatedly emphasized by 
Penrose \cite{Penrose:1965,PenroseRindler:book}. 
Exactly  this \defterm{directional radiation pattern},
i.e., the dependence of fields (with respect to appropriate tetrad) on the direction 
along which the null geodesic reaches a point on a spacelike $\scri^+$, 
is 
analyzed in the present work.

\sect{Spacetime infinity, fields and tetrads}{-2ex}{-1.7ex}

Following general formalism \cite{Penrose:1965,PenroseRindler:book}, 
a spacetime $\mfld$ with physical metric $\mtrc$ can be embedded into 
a larger \defterm{conformal manifold} $\cmfld$ with \defterm{conformal metric} $\cmtrc$
related to $\mtrc$~by ${\cmtrc=\om^2\mtrc}$.
Here, the conformal factor $\om$, negative in $\mfld$,
vanishes on the boundary  of $\mfld$ in $\cmfld$ called conformal infinity $\scri$. 
It is \defterm{spacelike} if the gradient $\grad\om$ on $\scri$ 
has timelike character. This may be either future infinity $\scri^+$ 
or past infinity $\scri^-$.  
Near $\scri^+$ we decompose $\cmtrc$ into a spatial 3-metric $\scrimtrc$
tangent to $\scri^+$ and a part orthogonal to~$\scri^+$:%
\begin{equation}\label{MtrcOnScri}
  \mtrc = \om^{-2} (-\clapse^2\,\grad\om{}^2+\scrimtrc)\period
\end{equation}                                                                             
The conformal lapse function $\clapse$ can be chosen                                      \pagebreak[1]
to be
constant on~$\scri^+$, e.g., equal to ${\scale=\sqrt{3/\Lambda}}$.
The form \eqref{MtrcOnScri} allows                                                        
us to define a timelike unit vector $\norm$ normal to $\scri^+\!$,%
\begin{equation}\label{NormVect}
  \norm^{\mu}=\om^{-1}\clapse\,\mtrc^{\mu\nu}\,\grad_\nu\om\period
\end{equation}

Next, we denote the vectors of an \defterm{orthonormal} tetrad  as
${\tG,\,\qG,\,\rG,\,\sG}$, where $\tG$ is a unit timelike
vector and the remaining three are unit spacelike vectors.
With this tetrad we associate a \defterm{null} tetrad
${\kG,\,\lG,\,\mG,\,\bG}$, such that
${\kG\spr\lG=-1}$, ${\mG\spr\bG=1}$,
\begin{equation}\label{NormNullTetr}
\begin{aligned}
  \kG &= \textstyle{\frac1{\sqrt{2}}} (\tG+\qG)\comma&
  \lG &= \textstyle{\frac1{\sqrt{2}}} (\tG-\qG)\commae\\
  \mG &= \textstyle{\frac1{\sqrt{2}}} (\rG-i\,\sG)\comma&
  \bG &= \textstyle{\frac1{\sqrt{2}}} (\rG+i\,\sG)\period
\end{aligned}
\end{equation}
Various specific tetrads introduced below will be
distinguished by an additional label in subscript.

As usual, we parametrize the Weyl tensor $\WT_{\alpha\beta\gamma\delta}$ 
(representing the gravitational field) by 5 complex coefficients
\begin{gather}
  \WTP{}{0} = \WT_{\alpha\beta\gamma\delta}\,
    \kG^\alpha\,\mG^\beta\,\kG^\gamma\,\mG^\delta\comma
  \WTP{}{1} = \WT_{\alpha\beta\gamma\delta}\,
    \kG^\alpha\,\lG^\beta\,\kG^\gamma\,\mG^\delta\commae\notag\\
  \WTP{}{2} = -\WT_{\alpha\beta\gamma\delta}\,
    \kG^\alpha\,\mG^\beta\,\lG^\gamma\,\bG^\delta\commae\label{PsiDef}\\
  \WTP{}{3} = \WT_{\alpha\beta\gamma\delta}\,
    \lG^\alpha\,\kG^\beta\,\lG^\gamma\,\bG^\delta\comma
  \WTP{}{4} = \WT_{\alpha\beta\gamma\delta}\,
    \lG^\alpha\,\bG^\beta\,\lG^\gamma\,\bG^\delta\commae\notag
\end{gather}
and the electromagnetic field $\EMF_{\alpha\beta}$ by three coefficients
\begin{equation}\label{PhiDef}
\begin{gathered}
  \EMP{}{0} = \EMF_{\alpha\beta}\,
    \kG^\alpha\,\mG^\beta\comma
  \EMP{}{2} = \EMF_{\alpha\beta}\,
    \bG^\alpha\,\lG^\beta\commae\\
  \EMP{}{1} = {\textstyle\frac12}\,\EMF_{\alpha\beta}\,
    \bigl(\kG^\alpha\,\lG^\beta-\mG^\alpha\,\bG^\beta\bigr)\period
\end{gathered}
\end{equation}
These simply transform under 
special Lorentz transformations --- null rotations with
$\kG$ or $\lG$ fixed, boosts in ${\kG\textdash\lG}$ plane, and spatial 
rotations in ${\mG\textdash\bG}$ plane. For instance, under a null
rotation with $\lG$ fixed, parametrized by~${K\!\in\!\complexn}$:
\begin{gather}
\begin{gathered}\label{lfixed}
  \lG = \lO \comma
  \kG = \kO + \bar K \mO + K \bO + K\bar K\, \lO\commae\\
  \mG = \mO + K\,\lO\comma
  \bG = \bO + \bar K\,\lO\commae
\end{gathered}\\
\begin{aligned}\label{lfixedfield}
  \WTP{}{0} &= K^4 \WTP{\refT}{4} + 4 K^3 \WTP{\refT}{3} +
    6 K^2 \WTP{\refT}{2} + 4 K\, \WTP{\refT}{1} + \WTP{\refT}{0}\commae\\
  \EMP{}{0} &= K^2 \EMP{\refT}{2} + 2 K\, \EMP{\refT}{1} + \EMP{\refT}{0}\period
\end{aligned}
\end{gather}
Similarly, null rotations with $\kG$ fixed can be parametrized by
$L\in\complexn$. For boosts in ${\kG\textdash\lG}$ plane, 
cf.\ Eqs.~\eqref{RotIntRel}, \eqref{boostfield}.
For details and notation see, e.g., Ref.~\cite{KrtousPodolsky:RABHDS,Krameretal:book}.

Our goal is to investigate the field components in  an appropriate interpretation 
tetrad parallelly transported along 
all null geodesics $\geod(\afp)$ which terminate 
at $\scri^+$ at a point $\scripoint$. 
A geodesic reaches $\scri^+$ at an infinite value of 
the affine parameter $\afp$. The conformal factor $\om$ 
and lapse $\clapse$ can be expanded along the geodesic in powers of~$1/\afp$:%
\begin{equation}\label{ExpAlongGeod}
  \om \lteq \om_* \afp^{-1} + \dots\comma
  \clapse \lteq \clapse_+ + \dots\period
\end{equation}
The value $\clapse_+=\clapse|_{\scripoint}$ 
is the same for all geodesics ending at point $\scripoint$. 
We require that the approach of geodesics to $\scri^+$ is \vague{comparable}, 
independent of their \emph{direction},
so we assume also $\om_*$ to be constant. 
This is equivalent to fixing the energy ${E_\refT=-\tens{p}\spr\norm}$                                  \nopagebreak[3]
(${\tens{p}=\frac{D\geod}{d\afp}}$ being 4-momentum)                                                    \nopagebreak[3] 
at a given small value of $\om$, i.e.,                                                                  \pagebreak[2] 
at given proximity from the conformal infinity \cite{KrtousPodolsky:RABHDS}.

To define an \defterm{interpretation null tetrad} ${\kI,\,\lI,\mI,\,\bI}$,
we have to specify it in 
a comparable way for all geodesics along different directions. 
The geodesics reach the same point $\scripoint$ and we prescribe its form there. 
We require the null vector $\kI$ to be proportional to the tangent vector of the geodesic, 
\begin{equation}\label{kIdef}
  \kI=\frac1{\sqrt{2}\clapse_+}\,\frac{D\geod}{d\afp}\spcpnct;
\end{equation}
the factor is again chosen independent of the direction.
The null vector $\lI$ is fixed by ${\kI\spr\lI=-1}$ and by the requirement that
the normal vector $\norm$ belongs to the $\kI\textdash\lI$ plane.
The vectors ${\mI,\,\bI}$ (or, ${\rI,\,\sI}$) cannot be specified canonically
--- they will be chosen by Eqs.~\eqref{RotIntRel},~\eqref{rotTrefTrs}.

Now, the projection of $\kI$ on the normal $\norm$ is
\begin{equation}\label{kIdotnO}
  -\kI\spr\norm\lteq{\textstyle\frac1{\sqrt{2}}}\,\afp^{-1}\commae
\end{equation}
so that, as ${\afp\to\infty}$, the interpretation tetrad is \vague{infinitely boosted} with respect to
an observer with 4-velocity $\norm$. To see this explicitly, we introduce an auxiliary 
tetrad ${\tR,\,\qR,\,\rR,\,\sR}$ 
adapted to the conformal infinity, ${\tR=\norm}$, with $\qR$ oriented along 
the spatial direction of the geodesic,
\begin{equation}\label{RotTetr}
  \qR\propto\kI^\scriproj=(\kI\spr\norm)\,\norm+\kI\commae
\end{equation}
and we choose the remaining spatial vectors ${\rR,\,\sR}$ 
to coincide with those of the interpretation tetrad. 
Using Eqs.~\eqref{kIdef}, \eqref{kIdotnO} and the definition of $\lI$, we get
\begin{equation}\label{RotIntRel}
\begin{gathered}
  \kI = B_\intT\; \kR = \afp^{-1}{\textstyle\frac1{\sqrt2}}\,(\norm+\qR)\comma\mI=\mR\commae\\
  \lI = B_\intT^{-1}\, \lR = \;\;\afp\;{\textstyle\frac1{\sqrt2}}\,(\norm-\qR)\comma\bI=\bR\commae
\end{gathered}
\end{equation}
$B_\intT=1/\afp$ being the boost parameter. Under such a boost the field components
\eqref{PsiDef}, \eqref{PhiDef} transform as \cite{Krameretal:book,KrtousPodolsky:RABHDS}
\begin{equation}\label{boostfield}
  \WTP{\intT}{j}=B_\intT^{2-j}\,\WTP{\rotT}{j}\comma
  \EMP{\intT}{j}=B_\intT^{1-j}\,\EMP{\rotT}{j}\period
\end{equation}
Together with behavior \eqref{fieldsnearscri} of the field components in a tetrad adapted to
$\scri^+$ we obtain the \emph{peeling-off property}.

\sect{Directional pattern of radiation}{-2.5ex}{-2ex}

\begin{figure}
\includegraphics{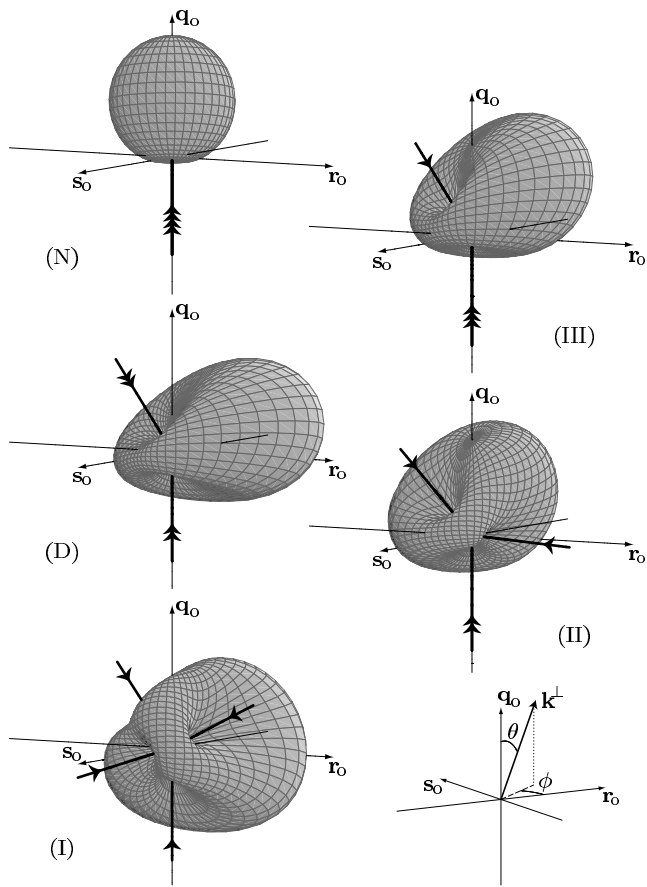}
\caption{\label{fig:dpr}%
Specific directional patterns of radiation for spacetimes of Petrov types 
\Ptype{N}, \Ptype{III}, \Ptype{D}, \Ptype{II} and \Ptype{I}. 
Directions in the diagrams represent all spatial directions tangent to $\scri^+$.
For each type, the radiative component $\abs{\WTP{\intT}{4}}$ along a null geodesic
is depicted in the corresponding spatial direction $\kG^\scriproj$
parametrized by spherical angles $\THT$, $\PHI$.
[Degenerate] principal null directions (PNDs) are indicated by [multiple] bold arrows. 
Thick lines represent spatial directions (opposite to PNDs) 
along which the radiation vanishes. 
}
\end{figure}

We shall now derive the directional dependence of radiation near a point $\scripoint$ at~$\scri^+$. 
It is necessary to parametrize the direction of the null geodesic reaching~$\scripoint$. This can be 
done with respect to a suitable \defterm{reference tetrad} ${\tO,\,\qO,\,\rO,\,\sO}$, 
with the time vector ${\tO}$ adapted to the conformal infinity, ${\tO=\norm}$, 
and spatial directions chosen arbitrarily. It is convenient 
to choose them in accordance with the spacetime geometry. 
Privileged choices will be discussed later (cf. Fig.~\ref{fig:dpr}).

The unit spatial direction $\qR$ of a general null geodesic near $\scri^+$ can be expressed in 
terms of \defterm{spherical angles} ${\THT,\,\PHI}$, with respect to the reference tetrad,
\begin{equation}\label{THTPHIdef}
  \qR = \cos\THT\;\qO + \sin\THT\cos\PHI\;\rO + \sin\THT\sin\PHI\;\sO\period 
\end{equation}
It is useful to introduce the \defterm{stereographic representation} 
of the angles ${\THT,\,\PHI}$,
\begin{equation}\label{Rdef}
  R = \tan(\THT/2)\,\exp(-i\PHI)\period
\end{equation}
Then the null rotation \eqref{lfixed} with ${K=R}$ transforms $\kO$ into $\kG$ with its spatial direction 
${\kG^\scriproj\!\propto\qR}$ specified by ${\THT,\,\PHI}$.

Now, the interpretation tetrad is related by boost \eqref{RotIntRel} to the tetrad 
${\tR,\,\qR,\,\rR,\,\sR}$, which is a spatial rotation of the reference tetrad. 
If we choose
\begin{equation}\label{rotTrefTrs}
\begin{aligned}
  \rR &= -\sin\THT\cos\PHI\,\bigl(\qO + \tan{\textstyle\frac\THT2}\,(\cos\PHI\;\rO+\sin\PHI\;\sO)\bigr) + \rO\commae\\
  \sR &= -\sin\THT\sin\PHI\;\bigl(\qO + \tan{\textstyle\frac\THT2}\,(\cos\PHI\;\rO+\sin\PHI\;\sO)\bigr) + \sO\commae
\end{aligned}
\end{equation}
the spatial rotation is a composition of the null rotations with $\lG$ 
fixed, $\kG$ fixed and the boost, given by parameters 
${K=R}$, ${L=-R/(1+\abs{R}^2)}$, ${B=(1+\abs{R}^2)^{-1}}$.
This decomposition into elementary Lorentz transformations enables 
us to calculate the field components in the interpretation 
tetrad. We start with $\WTP{\refT}{j}$, or $\EMP{\refT}{j}$ in 
the reference tetrad and we characterize these in terms of algebraically privileged 
\defterm{principal null directions} (PNDs). 

Principal null directions of the gravitational (or  electromagnetic) field are null directions $\kG$ such that 
${\WTP{}{0}=0}$ (or ${\EMP{}{0}=0}$)
in a null tetrad ${\kG,\,\lG,\,\mG,\,\bG}$ (a choice of ${\lG,\,\mG,\,\bG}$ is irrelevant). 
In the tetrad related to the reference tetrad by null rotation~\eqref{lfixed}, 
such a condition for $\WTP{}{0}$ takes the form
of a quartic (or quadratic for $\EMP{}{0}$) equation for $K$, 
cf.\ Eqs.~\eqref{lfixedfield}.
The roots ${K=R_n}$, ${n=1,2,3,4}$ (or ${K=R^\EM_n}$, ${n=1,2}$) of this equation thus 
parametrize PNDs $\kG_n$ (or $\kG^\EM_n$). As follows from the note 
after Eq.~\eqref{Rdef}, the angles ${\THT_n,\,\PHI_n}$ of these 
PNDs are related to $R_n$ exactly by Eq.~\eqref{Rdef}. 

In a generic situation 
we have ${\WTP{\refT}{4}\neq0}$ (or ${\EMP{\refT}{2}\neq0}$), and we can 
express the remaining components of the Weyl (or electromagnetic) tensor 
in terms of roots $R_n$ (or $R^\EM_n$)                                                                          \pagebreak[3]
\begin{gather}
\begin{aligned}
  \WTP{\refT}{3} &= -{\textstyle\frac14}\,\WTP{\refT}{4}\;(R_1+R_2+R_3+R_4)\commae\\
  \WTP{\refT}{2} &= \spcm{\textstyle\frac16}\,\WTP{\refT}{4}\;(R_1 R_2+R_1 R_3+R_1 R_4+\\
                 &\qquad\qquad\qquad\qquad R_2 R_3+R_2 R_4+R_3 R_4)\commae\\
  \WTP{\refT}{1} &= -{\textstyle\frac14}\,\WTP{\refT}{4}\;(R_1 R_2 R_3+R_1 R_2 R_4+\\
                 &\qquad\qquad\qquad\qquad R_1 R_3 R_4+R_2 R_3 R_4)\commae\\
  \WTP{\refT}{0} &= \spcm\spcm\WTP{\refT}{4}\; R_1 R_2 R_3 R_4\commae
\end{aligned}\\
  \EMP{\refT}{1} = -{\textstyle\frac12}\,\EMP{\refT}{2}\;(R^\EM_1+R^\EM_2)\comma
  \EMP{\refT}{0} = \EMP{\refT}{2}\,R^\EM_1R^\EM_2\period\quad
\end{gather}
Transforming these to tetrad ${\kR,\,\lR,\,\mR,\,\bR}$ we obtain
\begin{align}
\begin{split}
\WTP{\rotT}{4} &= \WTP{\refT}{4} \, \bigl(1+\abs{R}^2\bigr)^{-2}\\
&\quad\quad\times{\textstyle\bigl(1-\frac{R_1}{R\ant}\bigr)\bigl(1-\frac{R_2}{R\ant}\bigr)\bigl(1-\frac{R_3}{R\ant}\bigr)\bigl(1-\frac{R_4}{R\ant}\bigr)}\commae
\end{split}\label{WTProtT}\\
\EMP{\rotT}{2} &= \EMP{\refT}{2} \, \bigl(1+\abs{R}^2\bigr)^{-1}\;
{\textstyle \Bigl(1-\frac{R^\EM_1}{R\ant}\Bigr)\Bigl(1-\frac{R^\EM_2}{R\ant}\Bigr)}\period\label{EMProtT}
\end{align}
Here, the complex number $R\ant$,
\begin{equation}\label{antdef}
  R\ant = - {\bar R}^{-1} = -\cot(\THT/2)\,\exp(-i\PHI)\commae
\end{equation}
characterizes a spatial direction \defterm{opposite} to the direction given by $R$, 
i.e., the \defterm{antipodal} direction with ${\THT\ant=\pi-\THT}$ and ${\PHI\ant=\PHI+\pi}$.
Finally, we express the leading term of 
the field components in the interpretation tetrad. 
The freedom in the choice of the vectors ${\mI,\,\bI}$ changes just a phase of the field components, 
so only their modulus has a physical meaning.
It is also known \cite{Penrose:1965,PenroseRindler:book} that
as a consequence of field equations, field components in a reference tetrad
near $\scri^+$ behave as
\begin{equation}\label{fieldsnearscri}
  \WTP{\refT}{j}\lteq\WTP{\refT}{j}{}_*\;\afp^{-3}\comma
  \EMP{\refT}{j}\lteq\EMP{\refT}{j}{}_*\;\afp^{-2}\period
\end{equation}
Combining Eqs.~\eqref{boostfield} and \eqref{WTProtT}, \eqref{EMProtT}, we thus obtain
\begin{align}
\begin{split}
\abs{\WTP{\intT}{4}} &\lteq \abs{\WTP{\refT}{4}{}_*}\,\afp^{-1} \,\cos^4(\THT/2)\\
&\quad\qquad\times{\textstyle \bigabs{1-\frac{R_1}{R\ant}}\bigabs{1-\frac{R_2}{R\ant}}\bigabs{1-\frac{R_3}{R\ant}}\bigabs{1-\frac{R_4}{R\ant}}}\commae
\end{split}\label{WTPintT}\\
\abs{\EMP{\intT}{2}} &\lteq \abs{\EMP{\refT}{2}{}_*}\,\afp^{-1} \,
\cos^2(\THT/2)\;
{\textstyle \abs{1-\frac{R^\EM_1}{R\ant}}\abs{1-\frac{R^\EM_2}{R\ant}}}\period\label{EMPintT}
\end{align}

\sect{Discussion}{-3.25ex}{-1.5ex}

These expressions characterize the asymptotic
behavior of fields near de~Sitter-like infinity.  In a general
spacetime there are \emph{four} spatial directions along which
the radiative component of the gravitational field $\WTP{\intT}{4}$ vanishes,
namely directions ${R\ant = R_n\,}$, $n=1,2,3,4$
(or \emph{two} such directions for electromagnetic field $\EMP{\intT}{2}$).
In fact, their spatial parts $-\kG_n^\scriproj$ are exactly \emph{opposite to
the projections of the principal null directions $\kG_n^\scriproj$ onto~$\scri^+$}.

In algebraically special spacetimes, some
PNDs coincide and Eq.~\eqref{WTPintT} simplifies. 
Moreover, it is always possible to choose the \vague{canonical} 
reference tetrad: 
(i) the vector $\qO$ oriented along the spatial projection of the \emph{degenerate}
(multiple) PND, say $\kG_4^\scriproj$ (i.e.  ${\kO\propto\kG_4}$);
(ii) the ${\qO\textdash\rO}$ plane oriented so that it contains the spatial projection of
one of the remaining PNDs (for type~\Ptype{N} spacetimes this choice is
arbitrary). Using such a reference tetrad, the degenerate PND $\kG_4^\scriproj$ is 
given by ${\THT_4=0}$, i.e., ${R_4 = 0}$ (cf.\ Eq.~\eqref{Rdef}), 
whereas one of the remaining PNDs, say $\kG_1^\scriproj$,
has ${\PHI_1=0}$, i.e.\ 
${R_1 = \tan\frac{\THT_1}{2}}$ 
is real.

Thus, for Petrov type~\Ptype{N} spacetimes  (with quadruple PND)
${R_1=R_2=R_3=R_4=0}$,  so the asymptotic behavior of  gravitational  field \eqref{WTPintT} becomes 
\begin{equation}\label{WTdirN}
\abs{\WTP{\intT}{4}} = \abs{\WTP{\refT}{4}{}_*}\,\afp^{-1}\,\cos^4{\textstyle\frac\THT2}\period
\end{equation}
The corresponding directional pattern of radiation is illustrated
in Fig.~\ref{fig:dpr}\subfig{N}. It is axisymmetric, 
with maximum value at ${\THT=0}$ along the spatial
projection of the quadruple PND onto $\scri^+$.
Along the opposite direction, ${\THT=\pi}$, the field vanishes.

In the Petrov type~\Ptype{III} spacetimes,
${R_1=\tan\frac{\THT_1}{2}}$, 
${R_2=R_3=R_4=0}$, so \eqref{WTPintT} implies
\begin{equation}\label{WTdirIII}
\abs{\WTP{\intT}{4}}=\abs{\WTP{\refT}{4}{}_*}\,\afp^{-1}\,\cos^4\!{\textstyle\frac\THT2}\,
     \abs{1+\tan{\textstyle\frac{\THT_1}2}\,\tan{\textstyle\frac\THT2}\,e^{i\PHI}}\period
\end{equation}
The directional pattern of radiation is shown in Fig.~\ref{fig:dpr}\subfig{III}.                                \nopagebreak[3] 
The field vanishes along ${\THT=\pi}$ and along
${\THT=\pi-\THT_1}$, ${\PHI=\pi}$ which are spatial directions opposite to the PNDs.

The type~\Ptype{D} spacetimes admit two double degenerate PNDs,
${R_1=R_2=\tan\frac{\THT_1}{2}}$ 
and ${R_3=R_4=0}$.
The gravitational  field near $\scri^+$ thus takes the form%
\begin{equation}\label{WTdirD}
\abs{\WTP{\intT}{4}}=\abs{\WTP{\refT}{4}{}_*}\,\afp^{-1}\,\cos^4\!{\textstyle\frac\THT2}\,
    \abs{1+\tan{\textstyle\frac{\THT_1}2}\,\tan{\textstyle\frac\THT2}\,e^{i\PHI}}^2\commae
\end{equation}
with two planes of symmetry, see Fig.~\ref{fig:dpr}\subfig{D}. 
This directional dependence agrees with the radiation pattern for the $C$-metric
with ${\Lambda>0}$ derived recently \cite{KrtousPodolsky:RABHDS}.

For Petrov type~\Ptype{II} spacetimes only two PNDs coincide so that
${R_1=\tan\frac{\THT_1}{2}}$, ${R_2=\tan\frac{\THT_2}{2}\exp(-i\PHI_2)}$, 
${R_3=R_4=0}$.
The directional pattern of radiation is in Fig.~\ref{fig:dpr}\subfig{II}.
Finally, in case of algebraically general 
spacetimes one needs five real parameters
to characterize the directional dependence Fig.~\ref{fig:dpr}\subfig{I} of the gravitational field.

An analogous discussion can be presented for the electromagnetic field.
The square of expression \eqref{EMPintT} is, in fact,
the Poynting vector with respect to the interpretation tetrad,
${\abs{\EMS_\intT}\lteq\frac1{4\pi}\abs{\EMP{\intT}{2}}^2}$.
If the two PNDs coincide (${R^\EM_1=R^\EM_2=0}$)
the directional dependence of the Poynting vector at $\scri^+$ is 
the same as in Eq.~\eqref{WTdirN} (Fig.~\ref{fig:dpr}\subfig{N});
if they differ 
(${R^\EM_1=\tan\frac{\THT_1}{2}}$, 
${R^\EM_2=0}$), 
the  angular dependence of ${\abs{\EMS_\intT}}$ is given by Eq.~\eqref{WTdirD} (Fig.~\ref{fig:dpr}\subfig{D}).
The latter has already been  obtained for 
the test field of uniformly accelerated charges in de~Sitter spacetime \cite{BicakKrtous:FUACS}.

To summarize: it is well-known  that for \emph{null} $\scri^+$
the null direction $\lI$, which is complementary to the tangent vector $\kI$
of the null geodesic ${\geod(\eta)}$ reaching $\scri^+$, is
tangent to $\scri^+$ and does not depend on the choice of ${\geod(\eta)}$.
This is not the case when $\scri^+$ is spacelike. The radiation field
($\afp^{-1}$ term) is thus \vague{less invariant} \cite{PenroseRindler:book}. We have
shown that the dependence of this field on the choice of ${\geod(\eta)}$ has a universal
character that is determined by the algebraic (Petrov) type of the fields.
In particular, we have proved that the radiation  
vanishes along directions opposite to PNDs. 
In a \emph{generic} direction 
the radiative component of the fields
generated by \emph{any} source is \emph{nonvanishing}.
Thus, unlike in asymptotically flat spacetimes,
the absence of ${\afp^{-1}}$ component  cannot be used to
distinguish  nonradiative sources:
for a de~Sitter-like infinity the radiative component
reflects not only properties of the sources but also their \vague{kinematic} relation 
to an observer  at  infinity.
Intuitively, near spacelike $\scri^+$ our observer is,
in general, moving \vague{nonradially}  from sources
and thus measures  \vague{infinitely boosted} fields.
Some important questions, such as how energy is radiated away
in asymptotically de-Sitter spacetimes, still remain open.
Since vacuum energy seems to be dominant in our universe,
these appear to be of considerable interest.

\begin{acknowledgments}
The work was supported by GA\v{C}R 202/02/0735 and GAUK 166/2003. We thank J.~B.~Griffiths
for comments.
\end{acknowledgments}

\vspace*{-1ex}

\end{document}